\documentstyle[preprint,aps,tighten,eqsecnum,floats,graphics]{revtex}
\begin{document}
\draft
\widetext
\title{Frequency dependence of the photonic noise
spectrum in an absorbing or amplifying diffusive medium}
\author{
E. G. Mishchenko$^{1,2}$, M. Patra$^{1}$, and 
C. W. J. Beenakker$^{1}$}
\address{{}$^1$Instituut-Lorentz, Universiteit Leiden, P.O. Box 9506, 
2300 RA Leiden, The Netherlands}
\address{{}$^2$ L. D. Landau Institute for Theoretical Physics,
Russian Academy of Sciences, Kosygin 2, Moscow 117334, Russia}

\maketitle
\begin{abstract}
A theory is presented for the frequency dependence of the 
power spectrum of photon current fluctuations originating from
a disordered medium. Both the cases of an absorbing medium (``grey
body'') and of an amplifying medium (``random laser'') are considered
in a waveguide geometry. The semiclassical approach
(based on a Boltzmann-Langevin equation) is shown to be
in complete agreement with a fully quantum mechanical theory, 
provided that the effects of wave localization can be neglected.
The width of the peak in the power spectrum around zero frequency 
is much smaller than the inverse coherence time, characteristic
for black-body radiation. Simple expressions for the shape of this 
peak are obtained, in the absorbing case, for waveguide lengths large 
compared to the absorption length, and, in the amplifying case, close
to the laser threshold.
\medskip

\end{abstract}

\section{Introduction}

The noise power spectrum of a black body is frequency independent for
frequencies below the absorption band width. The inverse of the band width
is the coherence time $\tau_{\rm coh}$ of the radiation \cite{Lou83},
which for a black body is the longest relevant time scale --- hence the
white noise spectrum $P(\Omega)$ for $\Omega\lesssim 1/\tau_{\rm coh}$. In
a weakly absorbing, strongly scattering medium there appear two longer
time scales: The absorption time $\tau_{a}$ and the time $L^{2}/D$
it takes to diffuse (with diffusion constant $D$) through the medium
(of length $L$). As a consequence, $P(\Omega)$ for such a weakly-absorbing
medium (sometimes called a ``grey body'') starts to decay at much lower
frequencies than for a black body having the same coherence time.

Although there is by now a substantial literature on the theory
of grey-body radiation \cite{Bek94,Sch95,Lee95,Egb95,Buz98,Bee98},
the results have been limited to either the zero or high-frequency
limits of the noise spectrum (or, equivalently, to short or long
photodetection times). In the present work we remove this limitation,
by computing $P(\Omega)$ for a diffusive medium for arbitrary ratios
of $\Omega$, $1/\tau_{a}$, and $D/L^{2}$. We compare two different
approaches in a waveguide geometry: One which is fully quantum mechanical
(based on random-matrix theory \cite{Bee98,Pat99}) and another which is
semiclassical (based on a Boltzmann-Langevin equation \cite{MB}). Each
method has its advantages and disadvantages: The quantum theory includes
interference effects, which are ignored in the semiclassical
theory, but it is mathematically more involved. Complete agreement
between the two approaches is obtained in the limit that the waveguide
length $L$ is much smaller than the localization length (equal to the
mean free path times the number of propagating modes).

The results for absorbing media can be applied directly to linear
amplifiers, by formally changing the sign of the temperature and the
absorption time. Loudon and coworkers \cite{Jef93,Mat97} used this
relationship to calculate the noise power spectrum of a waveguide
without disorder. The generalization to a diffusive medium presented
here describes a random laser \cite{Cao99} below threshold.

The outline of this paper is as follows. We start with the semiclassical
approach, presenting a general solution of the Boltzmann-Langevin
equation in Sec.\ II and applying it to a waveguide geometry in Sec.\
III. The quantum mechanical approach is developed in Sec.\ IV. For the
quantum theory we need the correlator of reflection and transmission
matrices at different frequencies. These are calculated in the appendix,
using the random-matrix method of Ref.\ \cite{Bro98}. We discuss our
findings in Sec.\ V.

\section{Semiclassical theory}

\par Starting point of the semiclassical theory 
is the Boltzmann-Langevin
equation for photons of Ref.\ \cite{MB}.
We first consider an absorbing medium (in equilibrium
at temperature $T$), leaving the amplifying case for
the end of this section.
We make the diffusion approximation,
valid if the mean free path $l$ is the shortest length scale
in the system
(but still large compared to the wavelength).
The fluctuating
number density $n({\omega},{\bf r},t)$ and
current density ${\bf j}({\omega},{\bf r},t)$
of photons at frequency $\omega$, position
${\bf r}$, and time $t$ are related by \cite{MB}
\begin{eqnarray}
\label{asym}
&&{\bf j}= - D
\frac{\partial  n}{\partial {\bf r}}
+\mbox{\boldmath$\cal  L$}_1, \\
\label{sym}
&&\frac{\partial n}{\partial t}+
\frac{\partial }{\partial {\bf
r}}\cdot  {\bf j} = D \xi_a^{-2}(\rho f-n)
+{\cal L}_0.
\end{eqnarray}
Here $D=\case{1}{3}cl$ is the diffusion constant,
$\xi_a=\sqrt{D\tau_a}$ is the absorption length
(with $\tau_a$ the absorption time),
$\rho=4\pi\omega^2 (2\pi c)^{-3}$ is the density of
states (not counting polarizations), 
and $f=[\exp{(\hbar \omega/kT)}-1]^{-1}$ 
is the Bose-Einstein function.
We assume $\xi_a \gg l$.
The fluctuating source 
 terms
${\cal L}_0$ and $\mbox{\boldmath$\cal  L$}_1$ have zero mean
and correlators
\begin{mathletters}
\label{cor}
\begin{eqnarray}
\label{cor1}
&&\overline{{\cal L}_{0}(\omega,{\bf r},t) 
{\cal L}_{0}(\omega',{\bf r'},t') }
= \delta (\omega-\omega')
\delta (t-t') \delta ({\bf r} -{\bf r'})
D\xi_a^{-2}
(2f \bar{n}+\rho f+\bar{n}), \\
\label{cor2}
&&\overline{{\cal  L}_{1\alpha}(\omega,{\bf r},t) 
{\cal  L}_{1\beta}(\omega',{\bf r'},t') } 
= 2 \delta_{\alpha \beta} \delta (\omega-\omega')
\delta (t-t') \delta ({\bf r} -{\bf r'})
 D\bar{n}(1+\bar{n}/\rho). 
\end{eqnarray}
\end{mathletters}%
The cross-correlator of ${\cal L}_{0}$ 
and $\mbox{\boldmath$\cal  L$}_{1}$ is given in Ref.\ \cite{MB},
but will not be needed.
Combining Eqs.\ (\ref{asym}) and (\ref{sym}) we find equations
for the mean $\bar{n}$ and the fluctuations
$\delta n$ of the photon number density
$n=\bar{n}+\delta{n}$,
\begin{eqnarray}
\label{aver}
&&\mbox{} -\frac{1}{D}
\frac{\partial \bar{n}}{\partial t}+
\frac{\partial^2 \bar{n}}{\partial {\bf r}^2}- \frac{\bar{n}}{\xi_a^{2}}
=-\frac{\rho f}{\xi_a^{2}},\\
\label{flu}
&&\mbox{} -\frac{1}{D}
\frac{\partial \delta n}{\partial t}+
\frac{\partial^2 \delta n}{\partial {\bf r}^2}- \frac{ \delta n}{\xi_a^{2}}
=\frac{1}{D}\frac{\partial }{\partial {\bf
r}}\cdot \mbox{\boldmath$\cal  L$}_1 -\frac{ {\cal L}_0}{D}.
\end{eqnarray}
\begin{figure}[tb]
\label{1}
 \unitlength 1cm
 \begin{center}
 \begin{picture}(10,4.8)
 \put(-0.3,-3.4){\includegraphics{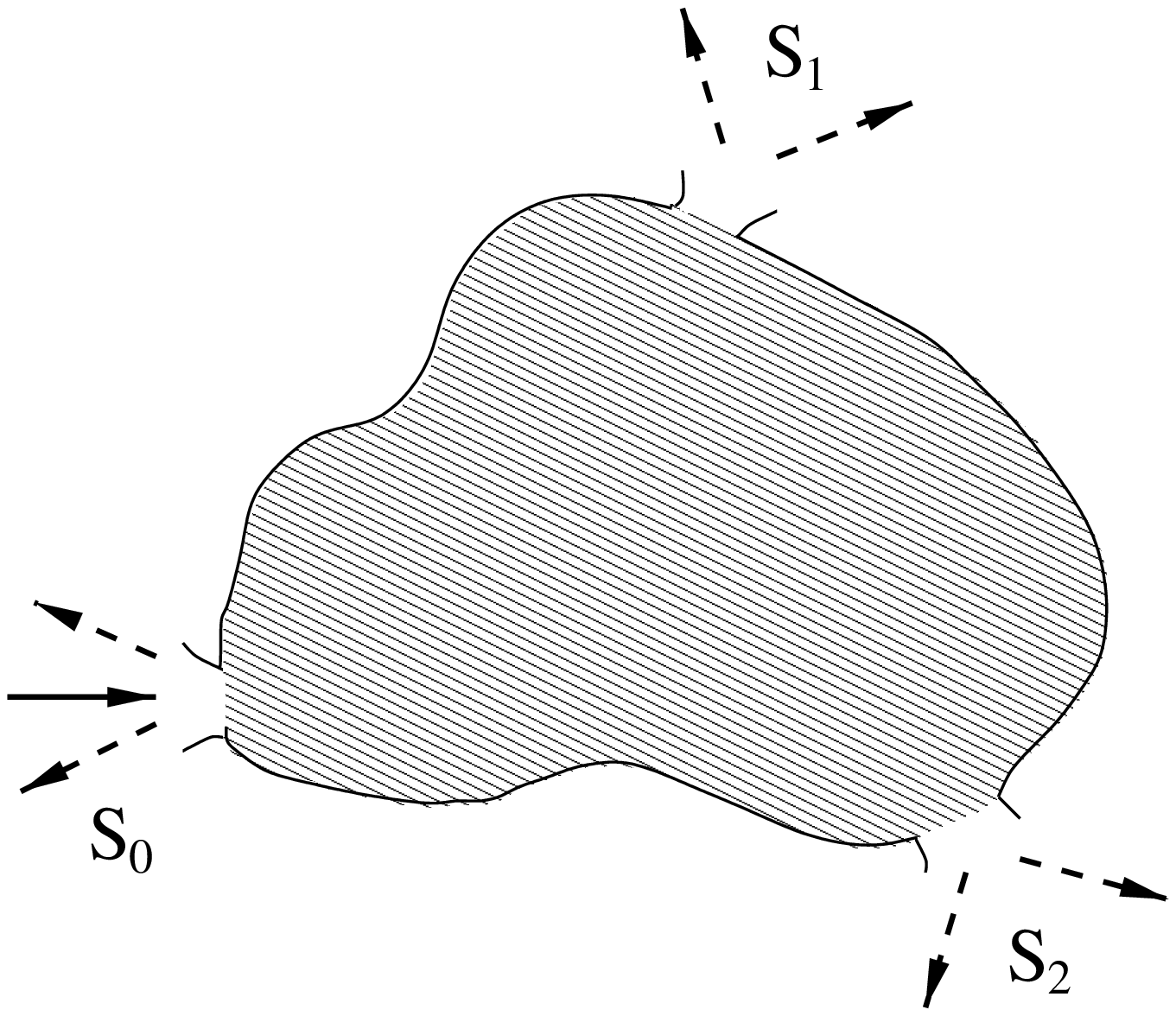}}
  \end{picture}
 \end{center}
  \caption[]{
Thermal radiation (solid arrow) is incident through
port $S_0$ on
an absorbing disordered medium (shaded). 
The outgoing
radiation (dashed arrows)
is absorbed by photodetectors.}
\end{figure}
\par We present a general solution
for the multiport geometry of Fig.\ 1.
Thermal 
radiation
is  incident through the
port $S_0$ and can leave the system 
via ports $S_0$, $S_1$, $S_2$, $\ldots$, 
where it is absorbed by photodetectors.
The corresponding boundary conditions are
$n(\omega, {\bf r},t) \vert_{{\bf r} \in S_p}= n_{\rm in}(\omega, t)
\delta_{p0}$.
We assume that the closed boundaries $\Sigma$ of the system
(with volume $V$)
are perfectly reflecting.
The separation of the ports is of order $L \gg l$.
In what follows we assume 
detection of outgoing radiation in a narrow frequency
interval $\delta \omega$ around $\omega$. 
We require that $\delta\omega$ is small both compared to
$\omega$ and to $1/\tau_{\rm coh}$.
To minimize the notations in this section we
 omit the frequency argument $\omega$
and use units in which $\delta \omega \equiv 1$.
(We will reinsert $\delta \omega$ in the next section.)
\par The Green function of the
differential equations (\ref{aver}) and (\ref{flu})
in the Fourier representation with respect to the time argument
satisfies
\begin{equation}
\label{gree}
\left(\frac{\partial^2}{\partial {\bf r}^2} -\xi_a^{-2}+\frac{i\Omega}{D}
\right)
 G({\bf r}, {\bf r'},\Omega)
=\delta({\bf r}-{\bf r'}).
\end{equation}
[Fourier transforms are defined as $f(\Omega)=\int_{-\infty}^{\infty} dt ~ e^{i\Omega t} f(t)$.]
For frequency resolved detection we require
$\Omega \ll \delta\omega$.
We impose the boundary conditions
\begin{mathletters}
\label{bound}
\begin{eqnarray}
\label{bound1}
 G({\bf r}, {\bf r'},\Omega)\arrowvert_{{\bf r} \in S_p} =0, ~ p=0,1,2, 
\ldots, \\
\label{bound2}
\mbox{\boldmath$\Sigma $}
\cdot
\frac{\partial G({\bf r}, {\bf r'},\Omega)}
{\partial {\bf r}}\arrowvert_{{\bf r} \in 
\Sigma}=0,
\end{eqnarray}
\end{mathletters}
where $\mbox{\boldmath$\Sigma $}$ denotes the outward normal 
direction to the surface $\Sigma$.
We consider separately the mean and the fluctuations
of the photon number and current densities.

\subsection{Mean solution}

The average photon density satisfying Eq.\ (\ref{aver})
can be expressed in Fourier 
representation in terms of the Green function (\ref{gree}),
\begin{equation}
\label{sol}
\bar{n}({\bf r},\Omega) = - 2\pi \rho f \xi_{a}^{-2}
 \delta(\Omega)
\int\limits_{V} d{\bf r'} ~G({\bf r}, {\bf r'},0)\nonumber\\
\mbox{} + \bar{n}_{\rm in}(\Omega)
\int\limits_{S_0} d {\bf S'} \cdot
\frac{\partial G({\bf r}, {\bf r'},\Omega)}
{\partial {\bf r'}}. 
\end{equation}
Substituting this formula into the
expression for the current (\ref{asym}) and integrating over the
area $S_p$ one obtains the mean outgoing current $\bar{I}_p$
through port $p \ne 0$,
\begin{eqnarray}
\label{th}
\bar{I}_{p} (\Omega)  = 2\pi \rho D f \xi_a^{-2}
 \delta (\Omega) 
\int\limits_{S_p} d{\bf S} \cdot \int\limits_V d{\bf r'} ~
\frac{\partial G({\bf r},{\bf r'},0)}{\partial {\bf r}} 
\nonumber\\
\mbox{} -
 D \bar{n}_{\rm in}(\Omega)
\int\limits_{S_p}
dS_{\alpha} \int\limits_{S_0} dS_{\beta}' 
~ \frac{\partial^2
G({\bf r}, {\bf r'},\Omega)}{\partial r_{\alpha}
\partial r'_{\beta}}.
\end{eqnarray}
(Summation over the repeating Greek 
indices is implied.)
The first term $\propto \delta (\Omega)$ 
is the time-independent mean thermal radiation
from the medium.
The second term is that part of the mean radiation
entering through port $0$ that leaves the medium through
one of the other ports. (The restriction to $p\ne 0$ is
not essential but simplifies the general formulas considerably,
so we will make this restriction in what follows.)

\subsection{Fluctuations}

\par The fluctuations in the number density
follow in a similar way from the Green function
and Eq.\ (\ref{flu}),
\begin{equation}
\label{denflu}
\delta n({\bf r},\Omega) = \frac{1}{D}
\int\limits_V d{\bf r'} ~
G({\bf r}, {\bf r'}, \Omega) \left(
\frac{\partial}{\partial {\bf r'}} \cdot
\mbox{\boldmath$\cal  L$}_1({\bf r'},\Omega)
-{\cal L}_0({\bf r'},\Omega) \right) 
\mbox{}+ \delta n_{\rm in}(\Omega) \int\limits_{S_0}
d{\bf S'} \cdot \frac{\partial G({\bf r}, {\bf r'},\Omega)}
{\partial {\bf r'}}.
\end{equation}
The fluctuation of the current density is then given by
Eq.\ (\ref{asym}),
\begin{eqnarray}
\label{deltacu}
\delta j_{\alpha}({\bf r},\Omega) = 
\int\limits_V d{\bf r'} \left( {G}_{\alpha \beta} 
({\bf r}, {\bf r'},\Omega) {\cal  L}_{1\beta} ({\bf r'},\Omega) 
+ 
\frac{\partial G ({\bf r}, {\bf r'},\Omega)}{\partial r_{\alpha}}
 {\cal L}_0({\bf r'},\Omega)\right)\nonumber\\
\mbox{} -D
\delta n_{\rm in}(\Omega) \int\limits_{S_0}
d S'_{\beta} ~ {G}_{\alpha \beta}({\bf r}, {\bf r'},\Omega).
\end{eqnarray}
We have defined
\begin{equation}
{G}_{\alpha \beta}({\bf r}, {\bf r'},\Omega)=\frac{\partial^2
G({\bf r}, {\bf r'},\Omega)}{\partial r_{\alpha}
\partial r'_{\beta}} +\delta_{\alpha \beta} 
\delta ({\bf r} -{\bf r'}).
\end{equation}
\par
We seek the 
correlator of the current fluctuations
\begin{equation}
C_{\alpha \beta} ({\bf r},\Omega; {\bf r'},\Omega')=
\overline{\delta j_{\alpha}({\bf r}, \Omega)
\delta j_{\beta}({\bf r'}, \Omega)}
\end{equation}
for ${\bf r} \in S_p$, ${\bf r'} \in S_{q}$ with $p,q \ne 0$.
With the help of Eqs.\ (\ref{cor}) and (\ref{deltacu}) 
it can be expressed as 
\begin{eqnarray}
\label{noisep}
&&C_{\alpha \beta}
({\bf r},\Omega; {\bf r'},\Omega')
= 
\frac{D}{\xi_a^2}
 \int\limits_V d{\bf r''} \frac{\partial G ({\bf r}, {\bf r''},\Omega)}{
\partial r_{\alpha}}\frac{\partial G ({\bf r'}, {\bf r''},
\Omega')}{
\partial r'_{\beta}} 
 \lbrack (2 f+1)
\bar{n}({\bf r''},\Omega+\Omega')+ \rho f \rbrack \nonumber\\
&&\mbox{+}
2 D \int\limits_V d{\bf r''}
{G}_{\alpha \gamma}
({\bf r}, {\bf r''},\Omega) {G}_{\beta \gamma}
({\bf r'}, {\bf r''},\Omega')  
\bigg[ \bar{n}({\bf r''},\Omega+\Omega')
+ \frac{1}{\rho}\int\limits \frac{d\Omega''}{2\pi} 
\bar{n}({\bf r''},\Omega+\Omega'')
\bar{n}({\bf r''},\Omega'-\Omega'') \bigg]. \nonumber\\
~
\end{eqnarray}
Following Ref.\ \cite{MB}, we have
neglected the term $\propto \delta n_{\rm in}$ in
Eq.\ (\ref{deltacu}) (smaller by a factor $l/L$)
and the cross-correlator
$\overline{{\cal L}_0\mbox{\boldmath$\cal  L$}_1}$ (smaller by a factor
$l/\xi_a$).
\par
 We now integrate ${\bf r}$ and ${\bf r'}$ over $S_p$ and
$S_{q}$ to obtain the correlator of the total currents through ports $p$ 
and $q$, 
\begin{equation}
C_{pq}(\Omega,\Omega') =
\int\limits_{S_p} dS_{\alpha} \int\limits_{S_q}  dS'_{\beta}
~ C_{\alpha \beta}
({\bf r},\Omega; {\bf r'},\Omega')
=C^{(1)}_{pq}(\Omega,\Omega')+
C^{(2)}_{pq}(\Omega,\Omega').
\end{equation}
The first term $C^{(1)}_{pq}$ contains the contribution
from the terms linear in the number density $\bar{n}$ 
in Eq.\  (\ref{noisep}). Performing integration  by parts 
and using Eqs.\ (\ref{gree})--(\ref{sol}) we find that this  
term vanishes for $p \ne q$. For $p=q$ it contains the mean 
current,
\begin{equation}
\label{shotn}
C^{(1)}_{pq}(\Omega,\Omega')=\delta_{pq} \bar{I}_p
(\Omega+\Omega').
\end{equation}
For a time-independent mean current $\bar{I}_p$
one has a white-noise spectrum 
$C^{(1)}_{pq}(\Omega,\Omega')=2\pi 
\delta_{pq}
\delta(\Omega+\Omega') \bar{I}_p.$
This is the usual shot noise, corresponding to
Poissonian statistics of the current fluctuations.
The second term  $C^{(2)}_{pq}$ 
describes the deviations from Poissonian
statistics. It arises from terms in
Eq.\ (\ref{noisep}) that are quadratic in $\bar{n}$.
Performing again  an integration by parts, one finds
\begin{eqnarray}
\label{excessn}
C^{(2)}_{pq}
(\Omega,\Omega')=\frac{2 D}{\rho}
\int\limits_{S_p} dS_{\alpha} \int\limits_{S_q}  dS'_{\beta}
\int\limits_{V}d{\bf r''}
\int\limits \frac{d\Omega''}{2\pi}
\frac{\partial \bar{n}({\bf r''},\Omega+\Omega'')}
{\partial r''_{\gamma}}
\frac{\partial \bar{n}({\bf r''},\Omega'-\Omega'')}
{\partial r''_{\gamma}}
\nonumber\\
\mbox{} \times
 \frac{\partial G ({\bf r}, {\bf r''},\Omega)}{
\partial r_{\alpha}}
\frac{\partial G ({\bf r'}, {\bf r''},\Omega')}{
\partial r'_{\beta}}.
\end{eqnarray}
Equation (\ref{excessn}) together with Eq.\
(\ref{sol}) is the result that we need for our
analysis of the frequency dependence of the
noise spectrum.

\subsection{Amplifying medium}

The extension of our general formulas to an amplifying medium
(in the linear regime below the laser threshold) is straightforward
\cite{MB}: We assume that the frequency $\omega$ at which we are detecting the radiation is close to the frequency
of an atomic transition with (on average) $N_{\rm upper}$ 
and $N_{\rm lower}$ atoms in the upper and lower state. Then the Bose-Einstein function can be replaced by the population inversion factor
$f=N_{\rm upper}(N_{\rm lower}-N_{\rm upper})^{-1}$.
This factor is negative in the amplifying case
(when $N_{\rm upper} > N_{\rm lower}$), 
with $f=-1$ for a complete
population inversion. (Equivalently, one can evaluate
$f$ at a negative temperature \cite{Mat97}, with $T \rightarrow 0^-$ for
complete inversion.) An amplifying medium has a negative absorption time 
$\tau_a= \xi_a^2/D$. We can account for this by
taking $\xi_a$ imaginary. With these two substitutions for $f$ and $\xi_a$
our formulas for an absorbing medium carry over to the amplifying
case.

\section{Waveguide geometry}

For the application of our general formulas we consider a waveguide 
geometry (see Fig.\ 2).
The waveguide has length $L$ and cross-sectional area
$A$, corresponding to
$N=\omega^2 A/4\pi c^2$ propagating modes 
(not counting polarizations) at frequency $\omega$.
We abbreviate $s=L/\xi_a$. We consider 
a stationary incident current $I_0=
\case{1}{4}cA\delta \omega\bar{n}_{\rm in}=(N\delta \omega/2\pi \rho)\bar{n}_{\rm in}$,
and calculate the noise power spectrum of the
transmitted current,
\begin{equation}
\label{powerd}
P(\Omega) = \int \limits_{-\infty}^{\infty}
dt ~ e^{i\Omega t} ~
\overline{\delta I(t) \delta I(0)}.
\end{equation}
In terms of the correlator of the previous section,
one has
 $C_{11} (\Omega,\Omega') =2\pi  P(\Omega) \delta(\Omega+\Omega')$.
\begin{figure}[tb]
\label{two}
  \unitlength 1cm
  \begin{center}
  \begin{picture}(8,2.2)
 \put(-0.3,-4.4){\includegraphics{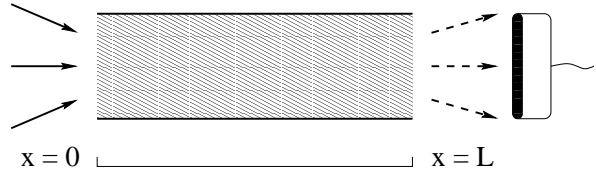}}
  \end{picture}
  \end{center}
  \caption[]{
Thermal radiation (solid arrows) is incident
on a  waveguide
containing an absorbing or amplifying disordered 
medium. The transmitted
radiation (dashed arrows) is absorbed by a photodetector.}
\end{figure}

\subsection{Absorbing medium}

\par We calculate the noise power from Eqs.\
(\ref{sol}) and (\ref{excessn}), using the
Green function
\begin{equation}
\label{gre}
G(x,x',\Omega)=-
\xi_a\frac{\sinh{[(x_</\xi_a)\sqrt{1-i\Omega \tau_a}]}
 \sinh{[(s-x_>/\xi_a)\sqrt{1-i\Omega \tau_a}]}}
{\sinh{[s\sqrt{1-i\Omega \tau_a}]}},
\end{equation}
where $x_<$ and $x_>$ are the smallest and largest of 
$x,x',$ respectively. 
The mean photon density is time independent. In Fourier
representation one has, from Eq.\ (\ref{sol}),
\begin{eqnarray}
\label{prof}
\bar{n}(x,\Omega)=  2\pi\delta(\Omega) \frac{\rho f}{\sinh{s}} 
\Bigl( \sinh{s}-\sinh{(x/\xi_a)}
-\sinh{(s-x/\xi_a)} \Bigr)
\nonumber\\  \mbox{}+2\pi\delta(\Omega)
\bar{n}_{\rm in}
\frac{\sinh{(s-x/\xi_a)}}{\sinh{s} }.
\end{eqnarray}
The  mean current $\bar{I}=\bar{I}_{\rm th}+\bar{I}_{\rm trans}$
is the sum of the thermal radiation 
from the medium 
\begin{equation}
\label{themean}
\bar{I}_{\rm th}=\frac{4Df}{c\xi_a}
(N\delta \omega/2\pi) \tanh{(s/2)}
\end{equation}
and the transmitted incident current
\begin{equation}
\bar{I}_{\rm trans}= \frac{4DI_0}{c\xi_a \sinh{s}}.
\end{equation}
\par Substitution of  Eqs.\
(\ref{gre}) and (\ref{prof}) 
into Eq.\ (\ref{excessn}) yields the super-Poissonian noise 
$P-\bar{I}$
as
a sum of three terms, 
$P-\bar{I}=P_{\rm th}+P_{\rm trans}+P_{\rm ex}$,
with
\begin{eqnarray}
\label{shum-th}
P_{\rm th}(\Omega) & = &
\frac{8Df^2}
{c\xi_a} (N\delta \omega/2\pi) \int\limits_0^{s}
ds'~\left( \frac{\cosh{(s-s')}-\cosh{s'}}{\sinh ~ s} \right)^2
K(s',s), \\
\label{shum-in}
P_{\rm trans}(\Omega) & = &
\frac{8 D{I}_{0}^2}{c\xi_a}
(2\pi/N\delta \omega) \int\limits_0^{s}
ds'~ \frac{\cosh^2{(s-s')}}{\sinh^2 ~ s} 
K(s',s),\\
\label{shum-ex}
P_{\rm ex}(\Omega) & = &
\frac{16Df{I}_{0}}
{c\xi_a} \int\limits_0^{s}
ds'~ \frac{[\cosh{s'}-\cosh{(s-s')}]\cosh{(s-s')}}{\sinh^2 ~ s} 
K(s',s).
\end{eqnarray}
We have defined
\begin{equation}
\label{ker}
K(s',s) = \left| \frac{\sinh 
(s'\sqrt{1-i\Omega\tau_a})}{\sinh (s\sqrt{1-i\Omega\tau_a})} 
\right|^2.
\end{equation}
The two terms  $P_{\rm trans}$ and $P_{\rm th}$ describe separately
the noise power of the transmitted incident current 
 and of the thermal current from the medium. 
The term $P_{\rm ex}$ is the excess
noise due to the beating of the incident radiation with the thermal
fluctuations from the medium.
\begin{figure}[tb]
  \unitlength 1cm
  \begin{center}
  \begin{picture}(8,13)
 \put(-6.2,-9.2){\includegraphics{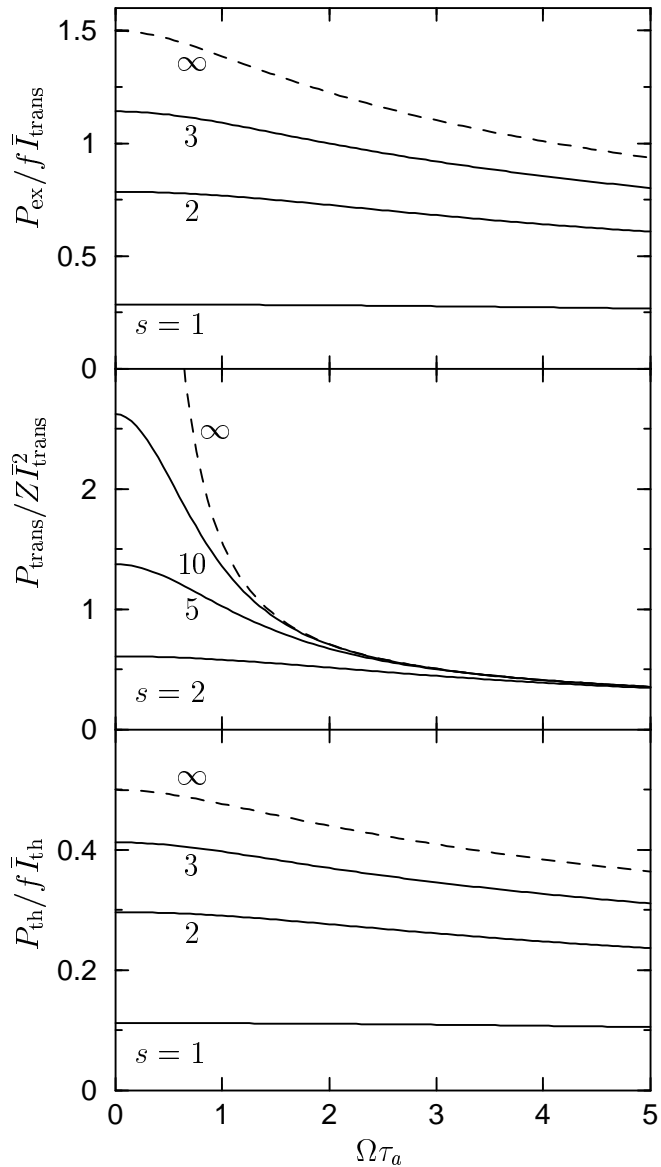}}
  \end{picture}
  \end{center}
  \caption[]{Frequency dependence 
of the three super-Poissonian contributions
to the noise power, $P-\bar{I}=P_{\rm th}+P_{\rm trans}+P_{\rm ex}$,
for different values of $s=L/\xi_a$ in an
absorbing waveguide. The solid curves are computed from
Eqs. (\ref{shum-th})--(\ref{shum-ex}),
the dashed curves are the large-$s$ asymptotes
(\ref{thbig})--(\ref{exbig}).
The parameter $Z$ is defined as
$Z=(c\xi_a/2D)(2\pi/N\delta\omega)$.}
\end{figure}
\par
The three contributions are plotted separately in Fig.\ 3.
For $L \gg \xi_a$ the frequency dependence simplifies to
\begin{eqnarray}
\label{thbig}
&& P_{\rm th}(\Omega) = 
\frac{f\bar{I}_{\rm th}}{1+x},\\
\label{trbig}
&&
P_{\rm trans}(\Omega) = \frac{c\xi_a \bar{I}_{\rm trans}^2}
{16D}(2\pi/N\delta\omega)\left(
\frac{1-e^{-2s(x-1)}}{x-1} +\frac{3x+2}{x^2+x}
\right),\\
\label{exbig}
&& P_{\rm ex}(\Omega) = 
f\bar{I}_{\rm trans} \frac{1+2x}{x+x^2},
\end{eqnarray}
where we have defined
\begin{equation}
x=\mbox{Re}~\sqrt{1-i\Omega\tau_a}=[\case{1}{2}(1+
\Omega^2\tau_{a}^2)^{1/2}+\case{1}{2}]^{1/2}.
\end{equation}
\par As discussed in Ref.\ \cite{MB} (for the zero-frequency case)
the result for $P_{\rm trans}$ requires that the incident radiation
is in a thermal state, at some temperature $T_0$.
(The quantity $f(\omega,T_0)=I_0(2\pi/N\delta \omega)$ is the corresponding value
of the Bose-Einstein function.)
There is no such requirement for $P_{\rm th}$ and $P_{\rm ex}$,
which are independent of the incident state.
For $T_0 \gg T$ we may generally neglect $P_{\rm th}$
and $P_{\rm ex}$ relative to $P_{\rm trans}$, so that
$P=\bar{I}_{\rm trans} +P_{\rm trans}$.
However, if the incident radiation is in a coherent state, then $P_{\rm trans}
\equiv 0$ and since  for sufficiently large $I_0$  we 
may neglect $P_{\rm th}$, we have in this case $P=\bar{I}_{\rm trans}+
P_{\rm ex}$. The contribution $P_{\rm th}$ is important mainly
in the absence of external illumination, when $P= \bar{I}_{\rm th}+
P_{\rm th}$.

\subsection{Amplifying medium}

The results for an amplifying medium are obtained by the substitution 
$\xi_a \rightarrow i\xi_a$, $f \rightarrow 
N_{\rm upper}(N_{\rm lower}-N_{\rm upper})^{-1}$, cf.\ Sec.\ IIC.
The frequency dependence of $P_{\rm th}, P_{\rm trans},$ and 
$P_{\rm ex}$ following from Eqs.\ (\ref{shum-th})--(\ref{shum-ex})
is plotted in Fig.\ 4 for lengths $L$ below
the laser threshold at $L=\pi \xi_a$.
\begin{figure}[tb]
\label{4}
  \unitlength 1cm
  \begin{center}
 \begin{picture}(8,13)
 \put(-6.2,-9.2){\includegraphics{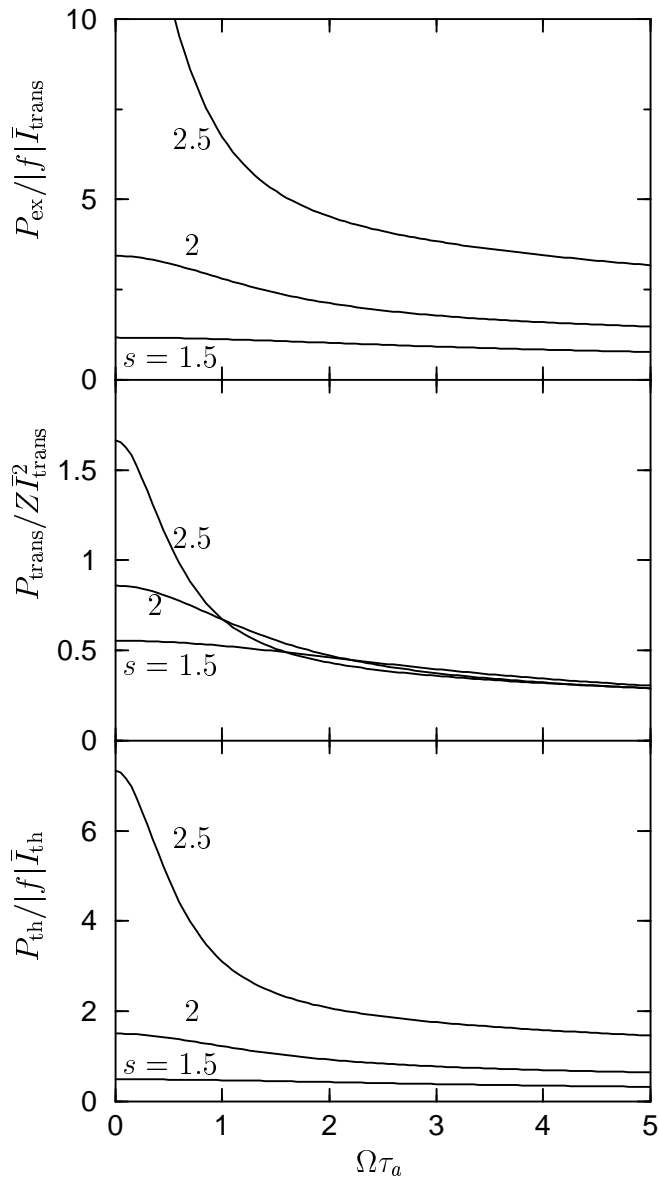}}
  \end{picture}
  \end{center}
  \caption[]{Same as Fig.\ 3, for the case of
an amplifying waveguide. The laser threshold occurs at
$s=\pi$.}
\end{figure}

\subsection{Cross-correlator}

In the absence of any incident radiation, the noise
$P= \bar{I}_{\rm th}+P_{\rm th}$
is due entirely to the thermal fluctuations in the medium. The 
current fluctuations at the two ends of the waveguide are
correlated, as measured by the cross-correlator

\begin{equation}
\label{power12}
P_{12}(\Omega) = \int \limits_{-\infty}^{\infty}
dt ~ e^{i\Omega t} ~
\overline{\delta I_1(t) \delta I_2(0)}.
\end{equation}
From Eqs.\ (\ref{excessn}), (\ref{gre}), and (\ref{prof})
we obtain 
\begin{eqnarray}
\label{mandel1}
P_{12}(\Omega) =  \frac{8D f^2}{c\xi_a} (N\delta\omega/2\pi)\int\limits_0^{s}
ds'~\left( \frac{\cosh{(s-s')}-\cosh{s'}}{\sinh ~ s} \right)^2 \nonumber\\
{}\times\frac{\sinh[ 
s'\sqrt{1-i\Omega\tau_a}]\sinh 
[(s-s')\sqrt{1+i\Omega\tau_a}]}{\vert \sinh [s\sqrt{1-i\Omega\tau_a}]
 \vert^2}.
\end{eqnarray}
The cross-correlator is plotted in Fig.\ 5 for both the 
absorbing  and amplifying cases. 
The outgoing currents at the two ends of the
waveguide are anti-correlated for $\Omega\tau_a \gg 1$.
\begin{figure}[tb]
\label{6}
  \unitlength 1cm
  \begin{center}
  \begin{picture}(8,9)
 \put(-6.5,-11.3){\includegraphics{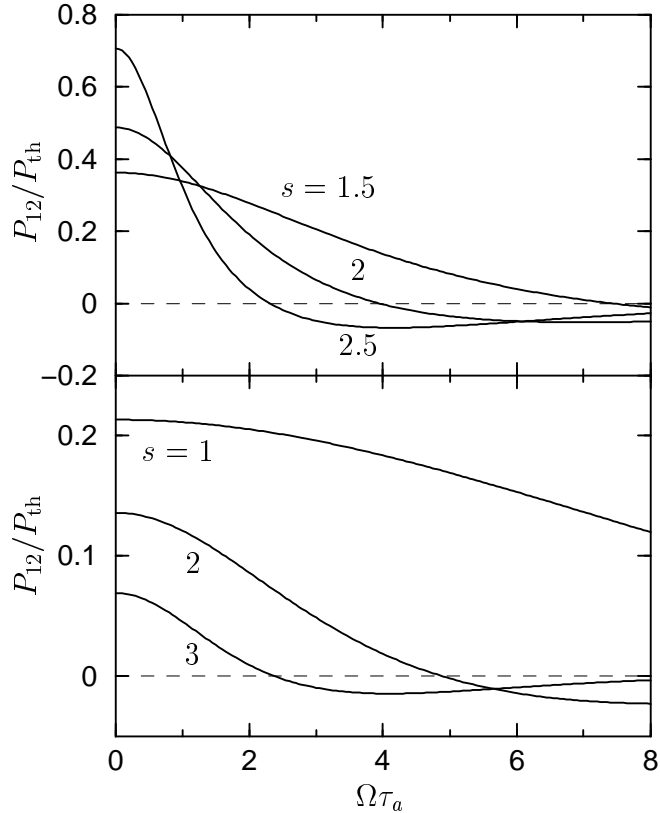}}
  \end{picture}
  \end{center}
  \caption[]{
Frequency dependence of the cross-correlator
of the outgoing current at the two ends of the
waveguide, in the absence of
any external illumination. Computed 
from Eq.\ (\ref{mandel1})  for the absorbing case
(lower panel) and amplifying case (upper panel).}
\end{figure}

\section{Comparison with quantum theory}

A fully quantum mechanical theory for the photocount distribution of
a disordered medium was developed in Refs.\ \cite{Bee98,Pat99}. In
this section we verify that it agrees with the semiclassical results
of the previous section. We 
consider the same system of Fig.\ 2, a disordered waveguide
with a photodetector at one end and a stationary current incident at
the other end. We assume that the incident current originates from a
thermal source at temperature $T_{0}$. The photocount distribution is the
distribution of the number of photons $n(t)$ counted (with unit quantum
efficiency) in the time interval $(0,t)$. Substitution of $I=dn/dt$
in the definition (\ref{powerd}) of the noise power $P(\Omega)$ leads to
a relation with the variance ${\rm Var}\,n(t)$ of the photocount,
\begin{mathletters}
\begin{eqnarray}
\label{pomega}
P(\Omega)=-\Omega^{2}\int\limits_{0}^{\infty}dt\,{\rm Var}\,n(t)\cos\Omega t,
\\
\mbox{Var}~n(t)=-\frac{2}{\pi}
\int\limits_0^{\infty} d\Omega ~ \Omega^{-2} P(\Omega) \left(
\cos{\Omega t}
-1\right).
\label{PVarrelation}
\end{eqnarray}
\end{mathletters}%
The variance can be separated into two terms, ${\rm Var}\,n(t)={\bar
n}(t)+\kappa(t)=t\bar{I}+\kappa(t)$, with $\kappa(t)$ the second
factorial cumulant. The term  $t\bar{I}$, substituted into Eq.\
(\ref{pomega}), gives the frequency-independent shot noise
contribution $\bar{I}$ to the power spectrum,
\begin{eqnarray}
\label{Pkapparelation}
P(\Omega)=\bar{I}-\Omega^{2}\int\limits
_{0}^{\infty}dt\,\kappa(t)\cos\Omega t.
\end{eqnarray}
The cumulant $\kappa=\kappa_{\rm trans}+\kappa_{\rm th}+\kappa_{\rm ex}$
contains separate contributions from the transmitted 
incident radiation and thermal
fluctuations in the medium, plus an excess contribution from the beating
of the two. These contributions have an exact representation in terms
of the $N\times N$ reflection and transmission matrices $r(\omega)$,
$t(\omega)$ of the waveguide \cite{Bee98,Pat99},
\begin{eqnarray}
\kappa_{\rm trans}(t)&=&\int\limits_{0}^{\infty}\frac{d\omega}{2\pi}
\int\limits_{0}^{\infty}\frac{d\omega'}{2\pi}
 L(\omega-\omega',t)\,f(\omega,T_{0})f(\omega',T_{0})
{\rm Tr}\,T(\omega)T(\omega'),\label{kinres}\\
\kappa_{\rm th}(t)&=&\int\limits\limits_{0}^{\infty}\frac{d\omega}{2\pi}
\int_{0}^{\infty}\frac{d\omega'}{2\pi}
 L(\omega-\omega',t)\,f(\omega,T)f(\omega',T)
{\rm Tr}\,Q(\omega)Q(\omega'),\label{kthres}\\
\kappa_{\rm ex}(t)&=&\int\limits_{0}^{\infty}\frac{d\omega}{2\pi}
\int\limits_{0}^{\infty}\frac{d\omega'}{2\pi}
 L(\omega-\omega',t)\,2f(\omega,T_{0})f(\omega',T)
{\rm Tr}\,T(\omega)Q(\omega'),\label{kexcres}
\end{eqnarray}
where we have defined
\begin{eqnarray}
 L(\omega,t)&=&\int\limits_{0}^{t}dt'\int\limits_{0}^{t}dt''\,
\exp[i\omega(t'-t'')]=2\omega^{-2}(1-\cos\omega t),\label{Ldef}\\
Q(\omega)&=&\openone-r(\omega)r^{\dagger}(\omega)-t(\omega)t^{\dagger}(\omega),\label{Qdef}\\
T(\omega)&=&t(\omega)t^{\dagger}(\omega).\label{Tdef}
\end{eqnarray}

Substitution into Eq.\ (\ref{Pkapparelation}) gives the corresponding
contributions to the noise power $P=\bar{I}+P_{\rm trans}+P_{\rm th}+P_{\rm ex}$,
\begin{eqnarray}
P_{\rm trans}(\Omega)&=&\frac{1}{2}\int\limits_{0}^{\infty}\frac{d\omega}{2\pi}
f(\omega,T_{0})f(\omega+\Omega,T_{0}){\rm Tr}\,T(\omega)T(\omega+\Omega)
+ \{\Omega\rightarrow -\Omega\},\label{Pinres}\\
P_{\rm th}(\Omega)&=&\frac{1}{2}\int\limits_{0}^{\infty}\frac{d\omega}{2\pi}
f(\omega,T)f(\omega+\Omega,T){\rm Tr}\,Q(\omega)Q(\omega+\Omega)
+ \{\Omega\rightarrow -\Omega\},\label{Pthres}\\
P_{\rm ex}(\Omega)&=&\frac{1}{2}\int\limits_{0}^{\infty}\frac{d\omega}{2\pi}
2f(\omega,T_{0})f(\omega+\Omega,T){\rm Tr}\,T(\omega)Q(\omega+\Omega)
+ \{\Omega\rightarrow -\Omega\}.\label{Pexcres}
\end{eqnarray}
As in the previous section, we assume a frequency-resolved measurement
in an interval $\delta\omega \ll \omega, 1/\tau_{\rm coh}$ with $\Omega\ll\delta\omega$. We
may then omit the integral over $\omega$ and approximate the argument
$\omega\pm\Omega$ in the functions $f$ by $\omega$. We take the ensemble
average $\langle\cdots\rangle$ of the noise power, in which case the
contributions from $\pm\Omega$ are the same. Finally, we insert the
incident current $I_{0}=f(\omega,T_{0})N\delta\omega/2\pi$, to arrive at
\begin{eqnarray}
P_{\rm trans}(\Omega)&=&(2\pi/N\delta\omega) I_{0}^{2}
\langle N^{-1}{\rm Tr}\,T(\omega)T(\omega+\Omega)\rangle,\label{Pinres2}\\
P_{\rm th}(\Omega)&=&(N\delta\omega/2\pi)f^{2}(\omega,T)
\langle N^{-1}{\rm Tr}\,Q(\omega)Q(\omega+\Omega)\rangle,\label{Pthres2}\\
P_{\rm ex}(\Omega)&=&2I_{0}f(\omega,T)
\langle N^{-1}{\rm Tr}\,T(\omega)Q(\omega+\Omega)\rangle.\label{Pexcres2}
\end{eqnarray}

It remains to evaluate the ensemble averages. This is done in the
appendix, by extending the approach of Ref.\ \cite{Bro98} to correlators
of reflection and transmission matrices at different frequencies. The
calculation applies to the diffusive regime that the length $L$ of
the waveguide is large compared to the mean free path $l$, but still
small compared to the localization length $Nl$. (The absorption length
$\xi_{a}$ is also assumed to be $\gg l$.) The results are
\begin{eqnarray}
\langle N^{-1}{\rm Tr}\,T(\omega)T(\omega+\Omega)\rangle&=&
\frac{8D}{c\xi_{ a}}\int\limits_{0}^{s}ds'\,K(s',s)\frac{\cosh^{2}(s-s')}{\sinh^{2}s},
\label{TrTTres}\\
\langle N^{-1}{\rm Tr}\,Q(\omega)Q(\omega+\Omega)\rangle&=&
\frac{8D}{c\xi_{a}}\int\limits_{0}^{s}ds'\,K(s',s)
\frac{[\cosh s'-\cosh(s-s')]^{2}}{\sinh^{2}s},
\label{TrQQres}\\
\langle N^{-1}{\rm Tr}\,T(\omega)Q(\omega+\Omega)\rangle&=&
\frac{8D}{c\xi_{a}}\int\limits_{0}^{s}ds'\,K(s',s)
\frac{\cosh(s-s')\cosh s'-\cosh^{2}(s-s')}{\sinh^{2}s},
\label{TrTQres}
\end{eqnarray}
where $s=L/\xi_{ a}$ and the kernel $K(s',s)$ is defined in Eq.\ (\ref{Kssdef}).
The combination of Eqs.\ (\ref{Pinres2})--(\ref{TrTQres}) agrees precisely with
the results (\ref{shum-th})--(\ref{shum-ex}) of the semiclassical theory. 
The quantum theory is more general than the semiclassical theory,
because it can describe the effects of wave localization. The method
of Ref.\ \cite{Bro98} gives corrections to the above results
in a power series in $L/Nl$. We will not pursue
this investigation here.

\section{Discussion}

We have presented a theory for the frequency dependence of the noise power spectrum $P(\Omega)$ in an absorbing or amplifying disordered
waveguide. The frequency dependence is governed by two time scales,
the absorption or amplification time $\tau_a$ and the
diffusion time $L^2/D$, both of which are assumed to be much greater than
the coherence time $\tau_{\rm coh}$. A simplified description is obtained, 
in the absorbing case, for lengths $L$ much greater than the absorption
length $\xi_a=\sqrt{D\tau_a}$, and, in the amplifying case, close
to the laser threshold at $L=\pi\xi_a$. We will discuss these two
cases separately.

\subsection{Absorbing medium}

The general formulas (\ref{shum-th})--(\ref{shum-ex})
for $P=\bar{I}+P_{\rm th}+P_{\rm trans}+P_{\rm ex}$ 
simplify for $L \gg \xi_a$ to Eqs.\ (\ref{thbig})--(\ref{exbig}). To characterize the frequency dependence we define the characteristic frequency $\Omega_c$ as
the frequency at which the super-Poissonian noise has
dropped by a factor of two:
\begin{equation}
P(\Omega_c) - \bar{I} = \case{1}{2} \left( P(0) -\bar{I} \right).
\end{equation}
In the absence of any external illumination ($I_0=0$)
we have, from Eq.\ (\ref{thbig}),
\begin{equation}
P=\bar{I}_{\rm th}\left(1+\frac{f}{1+x} \right),~~~~~ \bar{I}_{\rm th}=
\frac{4D f}{c\xi_a} (N\delta\omega/2\pi),
\end{equation}
with $x=\mbox{Re}~ \sqrt{1-i\Omega\tau_a}$, hence $\Omega_c 
=17/\tau_a$. If the illumination is in the coherent state from a laser,
then we have, from Eq.\ (\ref{exbig}),
\begin{equation}
P=\bar{I}_{\rm trans}\left(1+f\frac{1+2x}{x+x^2} 
\right),~~~~~ \bar{I}_{\rm trans}=
\frac{8 DI_0}{c \xi_a}e^{-s},
\end{equation}
here $\Omega_c = 9/\tau_a$.
In both these cases the diffusion time does not enter in the frequency
dependence. This is different for illumination by a thermal source
at temperature $T_0 $ much greater than the temperature of the 
medium. From Eq.\ (\ref{trbig}), with $f_0=f(\omega, T_0)$,
we then have
\begin{equation}
P_{\rm trans}(\Omega) = \bar{I}_{\rm trans}
\left(1+\frac{f_0}{2} e^{-s} \bigg[
\frac{1-e^{-2s(x-1)}}{x-1} +\frac{3x+2}{x^2+x} \bigg]
\right).
\end{equation}
The characteristic frequency $\Omega_c =
(64 D/L^2\tau_a^3)^{1/4}$ now contains both the
diffusion time and the absorption time.

\subsection{Amplifying medium}

In the amplifying case the noise power
becomes more and more strongly peaked near
zero frequency with increasing amplification. 
\begin{figure}[tb]
  \unitlength 1cm
  \begin{center}
  \begin{picture}(8,5.0)
 \put(-7.0,-12.7){\includegraphics{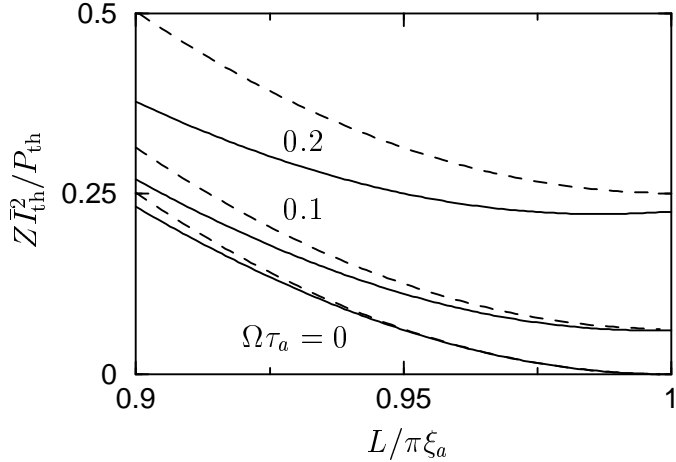}}
  \end{picture}
  \end{center}
  \caption[]{Ratio of  $\bar{I}_{\rm th}^2$ and
$P_{\rm th}$ in an amplifying waveguide
as a function of its length for different frequencies, computed
from Eqs.\ (\ref{themean}) and (\ref{shum-th}).
The approximation
(\ref{simple}) valid
near threshold for small frequencies is shown
dashed.}
\end{figure}
Close to the laser threshold at $s=\pi$
the frequency dependence
 of $P_{\rm th}$
for small frequencies $\Omega\tau_a \ll 1$ has the form
\begin{equation}
\label{simple}
P_{\rm th}=\frac{Z \bar{I}^2_{\rm th}}
{2\pi[\Omega^2\tau_a^2+4(1-s/\pi)^2]}, ~~~~
\bar{I}_{\rm th}=\frac{4f}{Z(\pi-s)}.
\end{equation}
Here again $Z=(c\xi_a/2D)(2\pi/N\delta\omega)$. 
Close to threshold the peak in the noise power spectrum
has a Lorentzian lineshape with half-width 
$\Omega_c = (2/\tau_a)(1-L/\pi\xi_a)$.
At the laser threshold both $P_{\rm th}$ and $\bar{I}_{\rm th}$
diverge, but the ratio 
$\bar{I}^2_{\rm th}/P_{\rm th}$ remains finite (see Fig.\ 6).
\par Finally, we note the fundamental difference
between the time scales appearing
in the noise spectrum for photons,
on the one hand, and electrons, on the other hand.
The absorption or amplification time $\tau_a$
obviously has no electronic analogue.
The diffusion time $L^2/D$ appears
in both contexts, however, the electronic noise
spectrum remains frequency independent for
$\Omega > D/L^2$ \cite{But}. The reason for the
difference is screening of electronic charge.
As a result the characteristic frequency scale
for electronic current fluctuations is
the inverse scattering time $D/l^2$, which is much
greater than the inverse
diffusion time $D/L^2$.

\acknowledgements
We thank P.\ W.\ Brouwer for advice concerning
the calculation in the appendix and Yu. V.\ Nazarov 
and M.\ P.\ van Exter for
useful discussions.
This research was supported by the ``Ne\-der\-land\-se or\-ga\-ni\-sa\-tie
voor We\-ten\-schap\-pe\-lijk On\-der\-zoek'' (NWO) and by the
``Stich\-ting voor Fun\-da\-men\-teel On\-der\-zoek der Ma\-te\-rie''
(FOM). E.\ G.\ M.\ also thanks the Russian Foundation for
Basic Research.
\appendix
\section{Correlators of reflection and transmission matrices}

To compute the noise power spectrum in the quantum mechanical approach of
Sec.\ V, we need the correlators of reflection and transmission
matrices $t(\omega_{\pm})$ and $r(\omega_{\pm})$ at two different
frequencies $\omega_{\pm}=\omega\pm\Omega/2$. (For $\Omega\ll\omega$ this
is the same as the correlator at frequencies $\omega$ and $\omega+\Omega$.)
We calculate these correlators for a waveguide geometry in the diffusive
regime, by extending the equal-frequency ($\Omega=0$) theory of Brouwer
\cite{Bro98}.

Upon attachment of a short segment of length $\delta L$ to one end of the
waveguide of length $L$, the
transmission and reflection matrices change according to
\begin{mathletters}
\label{deltart}
\begin{eqnarray}
t&\rightarrow&t_{\delta L}(1+r r_{\delta L})t,\label{deltarta}\\
r&\rightarrow&r_{\delta L}'+ t_{\delta L}(1+r r_{\delta L})r t_{\delta L}^{\rm T},
\label{deltartb}
\end{eqnarray}
\end{mathletters}
where the superscript ${\rm T}$ indicates the transpose of a
matrix. (Because of reciprocity the transmission matrix from left to
right equals the transpose of the transmission matrix from right to
left.) The transmission matrix $t_{\delta L}$ of the short segment at
frequency $\omega_{\pm}$ may be chosen proportional to the unit matrix,
\begin{equation}
t_{\delta L}=\left(1-\frac{\delta L}{2l'}-\frac{\delta L}{2c'\tau_{a}}
\pm\frac{i\Omega\delta L}{2c'}\right)\openone.\label{tdeltaLdef}
\end{equation}
 The mean free path $l'=4l/3$ 
and
the velocity $c'=c/2$ represent a weighted average over the $N$ transverse
modes in the waveguide.

Unitarity of the scattering matrix dictates that the reflection matrix
from the left of the short segment is related to the reflection matrix
from the right by $r'_{\delta L}=-r^{\dagger}_{\delta L}$. We abbreviate
$r_{\delta L}\equiv\delta r$. The matrix $\delta r$ is symmetric
(because of reciprocity), with zero mean and variance
\begin{equation}
\langle \delta r^{\vphantom{\ast}}_{kl}\delta r^{\ast}_{mn}\rangle=(N+1)^{-1}
(\delta_{km}\delta_{ln}+\delta_{kn}\delta_{lm})\delta L/l'.\label{deltartvar}
\end{equation} 
The resulting change in the matrix products $tt^{\dagger}$ and
$rr^{\dagger}$ is
\begin{mathletters}
\label{deltarrtt}
\begin{eqnarray}
tt^{\dagger}&\rightarrow&(1-\delta L/l'-\delta L/c'\tau_{a})tt^{\dagger}+
(r\delta rt)(r\delta rt)^{\dagger}\nonumber\\
&&\mbox{}+r\delta rtt^{\dagger}+(r\delta rtt^{\dagger})^{\dagger},
\label{deltarrtta}\\
rr^{\dagger}&\rightarrow&(1-2\delta L/l'-2\delta L/c'\tau_{a})rr^{\dagger}+
(r\delta rr)(r\delta rr)^{\dagger}+
\delta r^{\dagger}\delta r\nonumber\\
&&\mbox{}+r\delta rrr^{\dagger}+(r\delta rrr^{\dagger})^{\dagger}-
r\delta r-(r\delta r)^{\dagger}.\label{deltarrttb}
\end{eqnarray}
\end{mathletters}
The frequency $\Omega$ does not appear explicitly in these increments.

We define the following ensemble averages
\begin{eqnarray}
{\cal R}&=&\langle N^{-1}{\rm Tr}\,(\openone-rr^{\dagger})\rangle,\label{calRdef}\\
{\cal C}&=&\langle N^{-1}{\rm Tr}\,(\openone-r_{-}^{\vphantom{\dagger}}
r_{+}^{\dagger})\rangle,\label{calCdef}\\
{\cal T}&=&\langle N^{-1}{\rm Tr}\,tt^{\dagger}\rangle,\label{calTdef}
\end{eqnarray}
where $r,t$ are evaluated at frequency $\omega$ and $r_{\pm},t_{\pm}$
at frequency $\omega\pm\Omega/2$. Similarly, we define the correlators
\begin{eqnarray}
C_{rr}&=&\langle N^{-1}{\rm Tr}\,
(\openone-r_{-}^{\vphantom{\dagger}}r_{-}^{\dagger})
(\openone-r_{+}^{\vphantom{\dagger}}r_{+}^{\dagger})\rangle,\label{Crrdef}\\
C_{rt}&=&\langle N^{-1}{\rm Tr}\,
(\openone-r_{-}^{\vphantom{\dagger}}r_{-}^{\dagger})
t_{+}^{\vphantom{\dagger}}t_{+}^{\dagger}\rangle,\label{Crtdef}\\
C_{tt}&=&\langle N^{-1}{\rm Tr}\,
t_{-}^{\vphantom{\dagger}}t_{-}^{\dagger}
t_{+}^{\vphantom{\dagger}}t_{+}^{\dagger}\rangle.\label{Cttdef}
\end{eqnarray}
We will see that, in the diffusive regime, these 6 quantities satisfy a
coupled set of ordinary differential equations in $L$.

The diffusive regime corresponds to the large-$N$ limit, in which
the length $L$ of the waveguide is much less than the localization
length $Nl$. In this limit we may replace Eq. (\ref{deltartvar})
by $\langle\delta r^{\vphantom{\ast}}_{kl}\delta r^{\ast}_{mn}\rangle=(\delta
L/Nl')\delta_{km}\delta_{ln}$. In the large-$N$ limit we may also replace
averages of products of traces by products of averages of traces. From
Eq.\ (\ref{deltarrtt}) we thus obtain the diffential equations
\begin{eqnarray}
l'\frac{d{\cal R}}{dL}&=&2\gamma(1-{\cal R})-{\cal R}^{2},\label{dcalRdL}\\
l'\frac{d{\cal C}}{dL}&=&2\gamma(1+i\Omega\tau_{a})(1-{\cal C})
-{\cal C}^{2},\label{dcalCdL}\\
l'\frac{d{\cal T}}{dL}&=&-\gamma{\cal T}-{\cal RT},\label{dcalTdL}\\
l'\frac{dC_{rr}}{dL}&=&-(4\gamma+{\cal C}+{\cal C}^{\ast}+2{\cal R})C_{rr}
+2{\cal R}({\cal R}+2\gamma),\label{dCrrdL}\\
l'\frac{dC_{rt}}{dL}&=&-(3\gamma+{\cal C}+{\cal C}^{\ast}+{\cal R})C_{rt}-{\cal T}C_{rr}
+2({\cal R}+\gamma){\cal T},\label{dCrtdL}\\
l'\frac{dC_{tt}}{dL}&=&-(2\gamma+{\cal C}+{\cal C}^{\ast})C_{tt}-2{\cal T}C_{rt}
+2{\cal T}^{2},\label{dCttdL}
\end{eqnarray}
with the definition $\gamma=l'/c'\tau_{a}$. The initial conditions are
that each of these 6 quantities $\rightarrow 1$ for $L\rightarrow 0$.

This set of differential equations may be simplified further if we
assume, as we did in the semiclassical theory, that the mean free path
is small compared to both the absorption length and the length of the
waveguide. All 6 quantities (\ref{calRdef})--(\ref{Cttdef}) are of order
$\sqrt{\gamma}$, which is $\ll 1$ if $l'\ll c'\tau_{a}$, so that we
obtain in leading order
\begin{eqnarray}
l'\frac{d{\cal R}}{dL}&=&2\gamma-{\cal R}^{2},\label{dcalRdL2}\\
l'\frac{d{\cal C}}{dL}&=&2\gamma(1+i\Omega\tau_{a})-{\cal C}^{2},\label{dcalCdL2}\\
l'\frac{d{\cal T}}{dL}&=&-{\cal RT},\label{dcalTdL2}\\
l'\frac{dC_{rr}}{dL}&=&-({\cal C}+{\cal C}^{\ast}+2{\cal R})C_{rr}
+2{\cal R}^{2},\label{dCrrdL2}\\
l'\frac{dC_{rt}}{dL}&=&-({\cal C}+{\cal C}^{\ast}+{\cal R})C_{rt}-{\cal T}C_{rr}
+2{\cal R}{\cal T},\label{dCrtdL2}\\
l'\frac{dC_{tt}}{dL}&=&-({\cal C}+{\cal C}^{\ast})C_{tt}-2{\cal T}C_{rt}
+2{\cal T}^{2}.\label{dCttdL2}
\end{eqnarray}
As initial condition we should now take that the product of each quantity
with $L$ remains finite when $L\rightarrow 0$.

Although the differential equations are coupled, they may be solved
separately for ${\cal R}$, ${\cal C}$, ${\cal T}$, $C_{rr}$, $C_{rt}$,
$C_{tt}$, in that order. In terms of the rescaled length 
$s=(2\gamma)^{1/2}L/l'=L/\xi_{a}$, the results are
\begin{eqnarray}
{\cal R}&=&\frac{(2\gamma)^{1/2}}{\tanh s},\label{calRresult}\\
{\cal C}&=&\frac{(2\gamma)^{1/2}\sqrt{1+i\Omega\tau_{a}}}
{\tanh s\sqrt{1+i\Omega\tau_{ a}}},\label{calCresult}\\
{\cal T}&=&\frac{(2\gamma)^{1/2}}{\sinh s},\label{calTresult}\\
C_{rr}&=&\frac{(8\gamma)^{1/2}}{\sinh^{2}s}\int\limits_{0}^{s}ds'\,K(s',s)
\cosh^{2}s',\label{Crrresult}\\
C_{rt}&=&\frac{(8\gamma)^{1/2}}{\sinh^{2}s}\int\limits_{0}^{s}ds'\,K(s',s)
\cosh(s-s')\cosh s',\label{Crtresult}\\
C_{tt}&=&\frac{(8\gamma)^{1/2}}{\sinh^{2}s}\int\limits_{0}^{s}ds'\,K(s',s)
\cosh^{2}(s-s'),\label{Cttresult}
\end{eqnarray}
where the kernel $K$ is defined by
\begin{equation}
K(s',s)=\left|\sinh s'\sqrt{1+i\Omega\tau_{a}}\right|^{2}\,
\left|\sinh s\sqrt{1+i\Omega\tau_{ a}}\right|^{-2}.
\label{Kssdef}
\end{equation}
These are the expressions used in Sec.\ 4 (where we have also
substituted $\sqrt{2\gamma}=4D/c\xi_{a}$). The remaining integrals
over $s'$ may be done analytically, but the resulting expressions are
rather lengthy so we do not record them here.  For $\Omega=0$ our results
reduce to those of Brouwer \cite{Bro98} (up to a misprint in Eq.\ (13c)
of that paper, where the plus and minus signs in the expression between
brackets should be interchanged).

\end{document}